# Software Effort Estimation using parameter tuned Models

Akanksha Baghe Meemansa Rathod Pradeep Singh

Department of Computer Science and Engineering
National Institute of Technology Raipur
G. E. Road, Raipur – 492010, Chhattisgarh, India
psingh.cs@nitrr.ac.in ,abaghel48@gmail.com, 163meemansa@gmail.com

*Abstract*— Software estimation is one of the most important activities in the software project. The software effort estimation is required in the early stages of software life cycle. Project Failure is the major problem undergoing nowadays as seen by software project managers. The imprecision of the estimation is the reason for this problem. Assize of software size grows, it also makes a system complex, thus difficult to accurately predict the cost of software development process. The greatest pitfall of the software industry was the fast-changing nature of software development which has made it difficult to develop parametric models that yield high accuracy for software development in all domains. We need the development of useful models that accurately predict the cost of developing a software product. This study presents the novel analysis of various regression models with hyperparameter tuning to get the effective model. Nine different regression techniques are considered for model development.

## I. INTRODUCTION

Software development effort estimation is the process of predicting the most realistic amount of effort required to develop or maintain software based on incomplete, uncertain and noisy data. Software is the most expensive component in many computer based systems. A large amount of bugs creates huge differences between gain and loss during the estimation of effort [1].

Software effort estimation is way for finding the most practical use of effort required to maintain the software. It is evaluated in terms of person-month. It acts as input for planning of project. Using software effort estimation we can find resources that are required to complete the project on scheduled time. Software effort estimation plays important role in the completion of any project. Accurate estimations lead towards the completion of the project on the right time. To manage the resources for developing the software, reliable estimation is very necessary. In the field of software engineering, estimation of software development effort has always been a challenging topic. The most important thing in this field is its accuracy and reliability. There are various variables in the software effort estimation. These variables help in estimating the cost of the project. Instead of determining these variables at the end of the project, we should estimate them before the implementation of the project. The accurate effort estimation is very helpful for any of the organization and ongoing projects in that organization.

## II. LITERATURE RESEARCH

A number of software size and effort metrics have been identified in the literature. Putman's SLIM (Software Life Cycle Management) model incorporating software size and development time parameters compute a software effort estimation based on the Rayleigh function [2]. Albrecht first introduces a function points [3] methodology to calculate software size. Albrecht and Gaffney [4] then show the relationship between function points and development effort. Kemerer [5] evaluated four cost estimation models (SLIM, COCOMO, Function Points and ESTIMACS) with using a data set that covers 15 large completed data-processing projects. Matson et al. [6] developed effort estimation equations using function points data taking from 104 projects. Zheng et al. [7] propose a linear equation for software effort estimation based on Albrecht's function point. The System Evaluation and Estimation of Resources-Software Estimation Model (SEER-SEM) estimates the development effort as a function of three parameters: effective software size, effective technology and staffing complexity [8].

Boehm [9] introduces the first COCOMO model for software development effort estimation. The model estimates effort based on size of software and pre-determined constants. Boehm's intermediate COCOMO model computes software development effort as a function of estimated software size and a set of cost drivers that consists of product, hardware, personnel and product characteristics [10]. The formula uses different sets of coefficients when calculating program effort for organic, semi-detached and embedded software projects.

Further, Nassif et al. [11] presented a log-linear regression model based on the use case point model (UCP) to calculate the software effort based on use case diagrams.Sharma and Kushwaha [12] propose a measure for the estimation of software development effort on the basis of requirement based complexity. Olga Fedotova and Leonor Teixeria, Helena Alvelos [13] explained Software development organization that is used for the capability maturity model integrated. They described software development organization evaluates the effort used by the software based on the field of expert.

### III. DATA SET AND METHODOLOGY

In this paper, we have used eight benchmark datasets to assess software effort estimation of various machine learning algorithms. Datasets are Albrecht, China, Coc-nasa93, Cocomo81, Cocomonasa_2, Cocomonasa_v1, Cocomo-sdr, Desharnais, Kitchenham, Maxwell, UCP_Dataset. These datasets are retrieved from public data repository [14] [15]. The dataset used belongs to different developing environment. Datasets are divided into training and testing set. First, the regression model is made to learn the coefficients of the input and output from the training set and these coefficients are used to predict the value of the testing set. By comparing the values of estimated output and the actual output, error is calculated.

Table I. Description of Albrecht data

| Features | MIN | MAX | MEAN | Std Dev |
|---|---|---|---|---|
| Input count | 7 | 193 | 40.25 | 36.913 |
| Output count | 12 | 150 | 47.25 | 35.169 |
| Inquiry | 0 | 75 | 16.875 | 19.337 |
| File | 3 | 60 | 17.375 | 15.522 |
| FPAdj | 0.75 | 1.2 | 0.989 | 0.135 |
| RawFP | 189.52 | 1902 | 638.53 | 452.653 |
| AdjFP | 199 | 1902 | 658.875 | 492.204 |
| Effort | 0.5 | 105.2 | 21.875 | 28.417 |

Table II. Description of China data.

| Features | MIN | MAX | MEAN | Std Dev |
|---|---|---|---|---|
| AFP | 9 | 17518 | 486.857 | 1059.171 |
| Input count | 0 | 9404 | 167.0982 | 486.338 |
| Output count | 0 | 2455 | 113.6012 | 221.274 |
| Enquiry | 0 | 952 | 61.6012 | 105.4228 |
| File | 0 | 2955 | 91.234 | 210.270 |
| Interface | 0 | 1572 | 24.234 | 85.04 |
| Effort | 26 | 54620 | 3921.048 | 6480.855 |
| Duration | 1 | 84 | 8.1792 | 7.347 |

Table III. Statistical profile of NASA data.

| Features | Min | Max | MEAN | Std Dev |
|---|---|---|---|---|
| Rely | 0.75 | 1.4 | 1.07 | 0.16 |
| Data | 0.94 | 1.16 | 1 | 0.07 |
| Cplx | 0.7 | 1.65 | 1.14 | 0.17 |
| Stor | 1 | 1.56 | 1.13 | 0.18 |
| Time | 1 | 1.66 | 1.12 | 0.185 |
| Acap | 0.17 | 1.46 | 0.93 | 0.14 |
| Pcap | 0.7 | 1.42 | 0.92 | 0.13 |
| Pexp | 0.9 | 1.21 | 1 | 0.08 |
| Aexp | 0.82 | 1.29 | 0.93 | 0.08 |
| Tool | 0.78 | 1.17 | 0.99 | 0.09 |
| Sced | 1 | 1.23 | 1.04 | 0.05 |
| KLOC | 0.9 | 1153 | 86.8 | 148 |
| Effort | 5.9 | 11400 | 644 | 1444 |

Table IV. Description of Desharnais data

| Features | Min | Max | MEAN | Std Dev |
|---|---|---|---|---|
| Length | 1 | 39 | 11.72 | 7.40 |
| Transactions | 9 | 886 | 179.90 | 143.31 |
| Entities | 7 | 387 | 122.33 | 84.88 |
| PointAdjust | 73 | 1127 | 302.23 | 179.68 |
| Envergure | 5 | 52 | 27.63 | 10.59 |
| PointsNonAdjust | 62 | 1116 | 287.63 | 185.11 |
| Effort | 546 | 23940 | 5046.31 | 4418.77 |

Table V. Description of Kitchenham data

| Features | Min | Max | MEAN | Std Dev |
|---|---|---|---|---|
| Function Point | 15.36 | 18137.48 | 527.67 | 1521.99 |
| Effort | 220 | 113930 | 3113.12 | 9597 |

Table VI. Description of Maxwell data

| Features | Min | Max | MEAN | Std Dev |
|---|---|---|---|---|
| Duration | 4 | 54 | 17.21 | 10.65 |
| Size | 48 | 3643 | 673.31 | 784.08 |
| Time | 1 | 9 | 5.58 | 2.13 |
| Effort | 583 | 63694 | 8223.21 | 10499.90 |

Table VII. Description of UCP data.

| Features | MIN | MAX | MEAN | Std Dev |
|---|---|---|---|---|
| UAW | 6 | 19 | 10.4507 | 4.9879 |
| Simple UC | 0 | 20 | 2.69014 | 2.87646 |
| Average UC | 3 | 30 | 15.76056 | 5.37843 |
| Complex UC | 5 | 27 | 14.29577 | 4.422 |
| UUCW | 250 | 610 | 385.49295 | 88.4838 |
| Effort | 5775 | 7970 | 6561.2676 | 667.885 |

Table VIII. Description of Kemerer data

| Features | MIN | MAX | MEAN | Std Dev |
|---|---|---|---|---|
| Hardware | 1 | 6 | 2.333333 | 1.676163 |
| Duration | 5 | 31 | 14.26667 | 7.544787 |
| KSLOC | 39 | 450 | 186.5733 | 136.8174 |
| AdjFP | 99.9 | 2306.8 | 999.14 | 589.5921 |
| RAWFP | 97 | 2284 | 993.8667 | 597.4261 |
| EffortMM | 23.2 | 1107.31 | 219.2479 | 236.0554 |

Leave-one-out approach is used in the this study as performed [16]. Leave-one-out approach is the degenerate case of K-Fold Cross Validation, where K is chosen as the total number of examples. Dataset with N cases, perform N experiments and for each experiment use N-1 instances for training and the remaining example for testing. The error is estimated as the average error rate on test examples $E = \frac{1}{N}\sum_{i=1}^{N} E_i$

### A. Performance Measures

The following measures are used to estimate the capability and evaluating the performance, any regression model.

*1) Mean Magnitude of Relative Error (MMRE):* This is also known as Mean Absolute Relative Error [17] and is given by the equation

$$MMRE = \frac{1}{n}\sum_{i=1}^{n}\frac{|P_i - A_i|}{A_i}$$

where the variables denote the following:

$P_i$ : The predicted value of ith data point

$A_i$ : The absolute value of ith data point

$n$ : Number of data points

*2) Root Mean Squared Error (RMSE):* [18] This is another performance evaluation parameter and is represented by the following equation:

$$E = \sqrt{\frac{1}{n}\sum_{i=1}^{n}(P_i - A_i)^2}$$

where the variables denote the following:

$P_i$ : The predicted value of ith data point

$A_i$ : The absolute value of ith data point

$n$ : Number of data points

The ideal case is when E = 0. The range of E is [0, infinity). In this study, we have used RSME measure to compare the results.

Nine machine learning algorithms, namely Extreme Learning Machine(ELM), Linear Regression Model(LM),Classification and Regression Techniques(CART), Random Forest(RF), Partial Least Squares(PLS),Gaussian Process (GP), Linear Regression with Backwards Selection (LRBS),Bayesian Generalized Linear Model (BGLM) and Multivariate Adaptive Regression Splines (MARS) are used. The algorithms are selected as they belong from different categories.

### IV. OUTPUT AND RESULTS

This section represents the results of the comparison of Regression techniques using Regression Models which are applied on various datasets - Albrecht, China, Cocomo, Kemerer, Kitchenham, Desharnais, Maxwell and UCP. The reliability of the results is assessed using the LOOCV - Leave One Out Cross Validation approach. The results are provided in tables for every dataset with the RMSE value of each regression technique.

Table 9 shows the results of Albrecht, China, Cocomo81, Cocomonasa_2, Cococmonasa_v1, Nasa93, Cocomo-sdr, Kemerer, Kitchenham, Desharnais, Maxwell and UCP datasets and there RMSE value. Regression Technique, for which RMSE value is very low, is the best Regression Model for the Dataset. We have applied 9 regression techniques on each data set and the best technique is selected on the basis of minimum RMSE value.

Figure 1 shows the comparative results of regression technique for all the dataset on the basis of RMSE values. It can be observed from the figure that in the majority of the cases the results of extreme learning machine (ELM) are found comparable or even better than other regression techniques. We have found that the results have very low RMSE value so we can say that the results by using train datasets are very near to the actual effort value of the residuals.

For Albrecht PLS has outperformed and in the UCP Random Forest has performed better than other eight machine learning algorithm. In China dataset GP has performed better than the other algorithms used in the study but the RMSE is high. In case of Kitchenham MARS algorithms has out performed. ELM has outperformed in case of Maxwell dataset. For Desharnais data the ELM has performed as a second best learner and RF is the best learner out of nine learners. For Nasa_v1 Mars has performed better than the outer eight algorithms. The second last column of the table nine shows the average of RSME of the results for all the dataset for corresponding algorithms. The last column shows the rank on the basis average RSME. Partial Least Squares (PLS) are the best performer out of nine algorithms in average RMSE. The Extreme learning machine(ELM) is the second best performer in the basis of average RMSE.

### V. CONCLUSION

Software effort estimation is an important part of software development. Software is now more complex and its attributes are increasing due to this importance of research on estimation has been raised. In this work extensive experiments with eight benchmark dataset a with nine machine learning algorithms the effort estimation is performed. A benchmark comparative analysis has been performed. Simple, techniques like Partial Least Squares(PLS) and feedforward neural net based ELM performed better than other techniques. Selection of a proper estimation technique can have a significant impact on the performance. Also hyperparameter tunning plays a major role in the preparation of models using learning techniques.These results also indicate that machine learning techniques can make a valuable contribution to the set of software effort estimation techniques. The future work can include the study of new software effort estimation methods and models that help us to understand the effort estimation process of the software. The work can be done by selecting a combination of machine learning technique or transfer learning techniques which provides better and accurate results. Future research could be done to assess the cost benefit analysis of models to determine whether a given effort prediction model would be economically viable.

Table IX. Summary of RMSE values of all datasets

| Tech / Datasets | Albrecht | UCP | China | Kemerer | Kitchenham | Maxwell | Desharnais | Nasa_v1 | Avg | Rank |
|---|---|---|---|---|---|---|---|---|---|---|
| ELM | 16.6331 | 629.94 | 1093.894 | 235.6355 | 2128.626 | 5620.87 | 3220.64 | 266.4382 | 1651.585 | 2 |
| LM | 15.8045 | 164.958 | 1054.242 | 278.3396 | 153645.8 | 7029.67 | 3421.006 | 3194.315 | 21100.52 | 9 |
| CART | 22.1157 | 240.049 | 3547.825 | 272.2882 | 9162.325 | 8037.73 | 4064.736 | 385.1687 | 3216.53 | 8 |
| RF | 12.7009 | 44.8934 | 1384.26 | 234.8798 | 8462.33 | 6498.34 | 3182.199 | 309.9302 | 2516.192 | 6 |
| PLS | 10.4974 | 154.077 | 1074.498 | 237.1944 | 1980.67 | 5655.74 | 3260.64 | 246.7667 | 1577.51 | 1 |
| GP | 14.6588 | 164.425 | 1004.784 | 248.9476 | 2183.005 | 6745.83 | 3244.425 | 281.2831 | 1735.92 | 4 |
| LRBS | 17.5739 | 150.628 | 5954.227 | 251.5505 | 9635.872 | 5823.03 | 3033.109 | 512.3311 | 3172.29 | 7 |
| BGLM | 15.8513 | 164.958 | 1054.205 | 278.0983 | 2148.409 | 7014.36 | 3375.378 | 421.6358 | 1809.112 | 5 |
| MARS | 12.1336 | 48.2936 | 1133.382 | 278.0625 | 1815.024 | 6558.64 | 3749.41 | 243.2443 | 1729.774 | 3 |

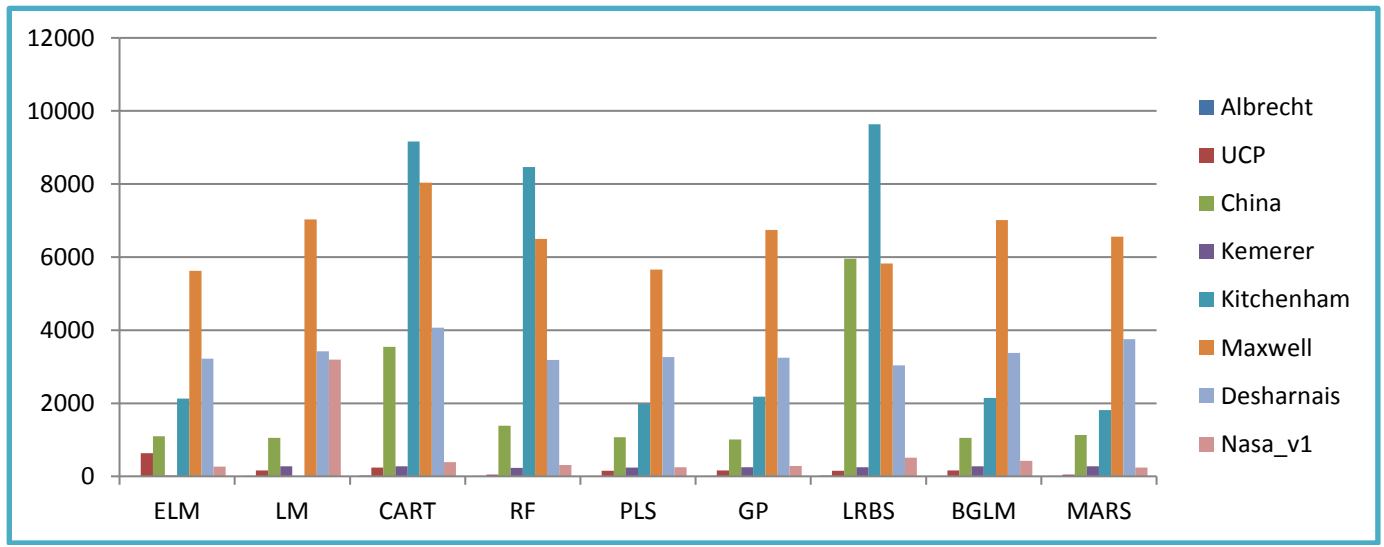

Figure 1. Bar Chart representation of table 8